\documentclass[twocolumn,showpacs,superscriptaddress,pra]{revtex4}
\usepackage{bm,graphicx,amsmath}

\begin{document}

\title{Quantum stability of Mott-insulator states of ultracold atoms in
optical resonators}

\author{Jonas Larson}
\affiliation{ICFO-Institut de Ci\`{e}ncies Fot\`{o}niques, E-08860
Castelldefels, Barcelona, Spain}

\author{Sonia Fern\'{a}ndez-Vidal}
\affiliation{ICFO-Institut de Ci\`{e}ncies Fot\`{o}niques, E-08860
Castelldefels, Barcelona, Spain}
\affiliation{Departament de
F\'{i}sica, Universitat Aut\`{o}noma de Barcelona, E-08193
Bellaterra, Spain}

\author{Giovanna Morigi}
\affiliation{Departament de F\'{i}sica, Universitat Aut\`{o}noma de
Barcelona, E-08193 Bellaterra, Spain}

\author{Maciej Lewenstein}
\affiliation{ICREA and ICFO-Institut de Ci\`{e}ncies Fot\`{o}niques,
E-08860 Castelldefels, Barcelona, Spain}
\affiliation{Institut
f\"{u}r Theoretische Physik, Universit\"{a}t Hannover, D-30167
Hannover, Germany}

\date{\today}

\begin{abstract} We investigate a paradigm example of cavity
quantum electrodynamics with many body systems: an ultracold
atomic gas inside a pumped optical resonator, confined by the
mechanical potential emerging from the cavity-field spatial mode
structure. When the optical potential is sufficiently deep, the
atomic gas is in the Mott-insulator state as in open space. Inside
the cavity, however, the potential depends on the atomic
distribution, which determines the refractive index of the medium,
thus altering the intracavity-field amplitude. We derive the
effective Bose-Hubbard model describing the physics of the system
in one dimension and study the crossover between the superfluid --
Mott-insulator quantum states. We predict the existence of
overlapping stability regions corresponding to competing
insulator-like states. Bistable behavior, controlled by the pump
intensity, is encountered in the vicinity of the shifted cavity
resonance. \end{abstract}


\maketitle

\section{Introduction} Cavity Quantum Electrodynamics (CQED)
\cite{walls-book,berman-book} has been a key area of quantum
optics since its early days  of  optical instabilities, such as
optical bistability \cite{carmichael}, till most recent CQED with
single atoms interacting with single or few photons
\cite{walther-review,haroche-book,schleich,kimble-review,rempe,grangier}.

In recent years considerable attention has been paid to a new
regime of CQED, which we term {\it CQED with many-body systems}.
These studies focus onto the mechanical effects of the resonator
field on the atomic motion, and on the non-linearity arising from
the interdependence between the cavity field and the atoms
dynamics. Following the theoretical prediction of
Ref.~\cite{Domokos}, signatures of self-organization have been
measured in the light scattered by laser-cooled atoms in a
transversally-pumped cavity \cite{black}. These structures and
their properties have been theoretically studied in detail in
Refs.~\cite{asboth,zippilli}. In different setups, Bragg
scattering of atomic structures inside optical resonators has been
experimentally investigated in Ref.~\cite{Zimmermann}.

While in  all the cases mentioned so far the atomic motion is
essentially classical, the stability and properties of these structures in the
quantum regime are still largely unexplored. This question acquires a
special relevance in view of the recent experimental progress of
CQED with {\it ultracold} atoms. In fact, strong atom-field
coupling between Bose-Einstein condensed atoms (BEC) and the mode
of a high-finesse optical cavity has been realized in the
experiments reported in
Refs.~\cite{reichel,reichel:CQED,zimmermannNew,Esslinger:BEC}.
Moreover, CQED techniques were used to measure pair correlations
in the atom laser \cite{esslinger}, and have been proposed for
characterizing quantum states of ultracold
matter~\cite{nature-ritsch,0610073,0702193}.

In~\cite{ourprl} we investigated the ground state of ultracold
atoms in the optical lattice formed by the interaction with the
cavity mode. This system combines CQED with the many-body physics
of {\it strongly correlated} ultracold atoms. In particular, the
non-linear dependence of the cavity field on the atomic motion
opens novel perspectives to the rich scenario of ultracold atomic
gases in optical lattices. In open space, in fact, these systems
offer the possibility to realize paradigmatic systems of quantum
many-body physics~\cite{bloch,ml}, such as various versions of
Hubbard models \cite{zoller}. A prominent example is the
Bose-Hubbard model \cite{fisher}, exhibiting the superfluid (SF)
-- Mott insulator (MI) quantum phase transition \cite{sachdev},
whose realization with ultracold atoms was proposed in
Ref.~\cite{jaksch}, and demonstrated in Ref.~\cite{blochex}.
In~\cite{ourprl} we addressed the question whether and how this
transition is modified when the atoms are inside a resonator,
where the optical lattice due to the intracavity-field depends on
the atomic density.

In this article we report the details of the derivations presented
in ~\cite{ourprl} and extend them to novel regimes. The system we
consider consists of ultracold atoms inside a resonator, which is
driven by a laser. Due to the strong coupling between cavity and
atomic degrees of freedom, the atoms shift the cavity resonance,
thus modifying the intracavity field intensity. This in turn
determines the depth of the cavity potential. At ultralow
temperatures we assume that the atoms occupy only states of the
lowest band of the periodic optical potential. In this regime, we
present the detailed derivation of the Bose-Hubbard model for
atoms in the one-dimensional potential of an optical resonator,
which complements and extends the derivation for few atoms by
Maschler and Ritsch in Ref.~\cite{Maschler,0611690} to large
numbers of atoms and which is valid in an appropriate
thermodynamic limit. Using this model we study the  SF -- MI
crossover as a function of the system parameters: the chemical
potential $\mu$, the pump strength and frequency, and the atomic
density. Assuming the tight-binding regime, we may describe the MI
states using Wannier functions \cite{mermin}, whose form is
determined by the optical potential. The Wannier functions are
used then to calculate the coefficients in the Bose-Hubbard model,
as in standard textbooks.  We determine the boundaries of the MI
states, using the {\it strong coupling expansion} of
Ref.~\cite{monien}, which is a quite accurate method for the
calculation of the phase-diagram of the BH model in 1D~\cite{ml}.

It must be stressed that the derivation of the Bose-Hubbard model in
the cavity does involve certain novel aspects. Namely, the periodic
optical potential depends functionally on the atomic density, and
hence on the Wannier functions. The problem is hence highly
non-linear: the coefficients of the equations determining the
Wannier functions depend functionally on the Wannier functions
themselves, and the latter have thus to be determined
self-consistently. This property has also physical implications. In
fact, since in our system the coefficients of the Hubbard model
depend on the atomic density, consequently, the diagrams in the
$\mu$-$t$ plane, where $t$ is the tunneling energy, exhibit
overlapping, competing Mott states, that may even consist of
disconnected regions for a wide range of parameters. In the vicinity
of the shifted cavity resonance in the strong coupling regime, the
situation is even more complex: one encounters there also dispersive
bistable behavior~\cite{carmichael}. Thus, from the quantum optical
perspective, this paper investigates stability of Mott-like phases,
i.e., insulator-like states, in an optical resonator.

This article is organized as follows. In Sec.~\ref{Sec:Model} we
present our model that describes a system of two level atoms
confined along the axis of an
optical cavity. In Sec.~\ref{Sec:Model:A} we introduce the
single-atom Hamiltonian. The many-body dynamics including the
quantum noise is introduced in Sec.~\ref{Sec:Many-Body}.
Section~\ref{Sec:Model:C} is devoted to the physical discussion of
the role of the various physical parameters on the system
dynamics. The effective Bose-Hubbard Hamiltonian is derived in
Sec.~\ref{Sec:BH}. In Sec.~\ref{Sec:GS} we discuss the
ground-state properties of our model. We use the Gaussian
approximation for the Wannier wave functions, checking carefully
its validity. Within the Gaussian approximation and the strong
coupling expansion method of H. Monien \cite{monien} the stability
regions for the Mott states are obtained analytically, up to the
solution of the non-linear self-consistent equation for the width
of the Wannier functions. Numerical results are reported and their
physical meaning is discussed in Sec.~\ref{Sec:Num}. The
validity of the approximations is addressed in Sec.~\ref{Sec:Approx}.
We conclude in Sec.~\ref{Sec:Conclusions}, while in the appendices
the details of the derivation of the Bose-Hubbard Hamiltonian and of
the strong coupling expansion method are reported.

\section{The model}
\label{Sec:Model}

In this section we generalize the quantum optical model of a single
atom inside a cavity to the many-body case, considering particle
collisions at ultralow temperatures and quantum statistics. At this
purpose, we first introduce the single-atom dynamics, write then the
Hamiltonian in second quantization, introducing atom-atom
collisions, and discuss the basic properties.

\subsection{Single-particle dynamics}
\label{Sec:Model:A}

\label{Sec:JC} We consider a single atom of mass $M$ inside a
cavity. The atomic dipole transition at frequency $\omega_0$,
between the ground state $|g\rangle$ and the excited state
$|e\rangle$, couples quasi-resonantly with an optical mode of the
resonator at frequency $\omega_c$, wave vector $k$ and
position-dependent coupling strength $$g(x)=g_0\cos (kx),$$ $g_0$
being the vacuum Rabi frequency. The resonator is driven by a
classical field of amplitude $\eta$ and oscillating at frequency
$\omega_p$. We consider the atomic motion along the cavity axis,
which coincides here with the $x$-axis, and assume tight
confinement along the radial plane so that the transverse motion
can be considered frozen out. Atomic center-of-mass position and
momentum operators are $x$ and $p$, fulfilling the uncertainty
relation $[x,p]={\rm i}\hbar$. In the reference frame oscillating
at the frequency $\omega_p$ of the pump field, the
normally-ordered Hamiltonian describing the coherent dynamics of
the atomic and cavity-mode states reads \begin{eqnarray}
\label{eq:JC}
H_{JC}&=&\frac{p^{2}}{2m}-\hbar \Delta_{a}\sigma^{\dagger}\sigma-\hbar \Delta_{c}a^{\dag}a\nonumber \\
&&-{\rm i}\hbar g_0\cos\left(x\right) \left( \sigma^{\dagger}a -
a^{\dag} \sigma \right)-{\rm i}\hbar\eta \left(a-a^{\dag}\right),
\end{eqnarray} where $\Delta_{a}= \omega_{p}-\omega_{a}$ and
$\Delta_{c} = \omega_{p}- \omega_{c}$ are the pump-atom and
pump-cavity detunings, $a$ ($a^{\dag}$) the annihilation
(creation) operator of a cavity photon at frequency $\omega_{c}$,
fulfilling the commutation relation $\left[a,a^{\dag}\right]=1$,
and $\sigma=|g\rangle\langle e|$,
$\sigma^{\dagger}=|e\rangle\langle g|$ are the dipole lowering and
raising operators. Spontaneous emission of the atomic dipole at
rate $\gamma$ and cavity losses at rate $\kappa$ are described
within the quantum Langevin equations formalism, such that the
quantum Heisenberg-Langevin equations for the dipole and cavity
operators read~\cite{walls-book} \begin{eqnarray}
\dot{a}(t)&=&(i\Delta_c-\kappa) a(t)+g(x)\sigma(t)+\eta
+\sqrt{2\kappa}a^{in}(t), \label{QLE:a}\\
\dot{\sigma}(t)&=&\left[i\Delta_a -\frac{\gamma}{2}\right] \sigma(t)+g(x)\sigma_za +\sqrt{\gamma}\sigma_zf^{in}(t), \label{penlastqle}\\
\dot{\sigma}_z(t)&=&
2g(x)\left[\sigma^{\dagger}a+a^{\dagger}\sigma\right]-\gamma
\left(\sigma_z(t)+1\right)/2\nonumber\\
& &
+2\sqrt{\gamma}\left(\sigma^{\dagger}f^{in}+f^{in\dagger}\sigma\right),
\label{QLE:z}
\end{eqnarray}
where
$\sigma_z=\sigma^{\dagger}\sigma-\sigma \sigma^{\dagger}$ and
$a^{in}$, $f^{in}$ are the input noise operators, whose mean value
vanishes and which are $\delta-$correlated in time, namely,
\begin{eqnarray}
&&\langle a^{in}(t)a^{in}(t')^{\dagger}\rangle = \delta(t-t'),\label{a:noise}\\
&&\langle f^{in}(t)f^{in}(t')^{\dagger}\rangle = \delta(t-t').
\end{eqnarray}
At large atom-pump detuning the adiabatic
elimination of the excited atomic state can be performed. Assuming
that the changes in the atomic position are negligible during the
time scale in which the atom reaches its internal steady state,
namely when $k_B T\ll \hbar|\Delta_a|$, we solve the
Heisenberg-Langevin equations at a fixed value of the atomic
position $x$. Hence, for $|\Delta_a|\gg g_0\sqrt{\langle
n\rangle},\gamma, |\Delta_c|$, we set $ \sigma_{z}(t)\approx-1$ in
the equations, and obtain $\sigma^{\dag}\approx{\rm
i}g(x)a^{\dag}/\Delta_{a}$. After tracing out the internal degrees
of freedom, the single-particle Hamiltonian for cavity and atomic
center-of-mass degrees of freedom reads
\begin{equation}\label{singleham}
H_{0}=\frac{\hat{p}^{2}}{2m}+\hbar\left[U_{0}\cos^{2}\left(k\hat{x}\right)-\Delta_{c}\right]\hat{a}^{\dag}\hat{a}
-i\hbar\eta\left(\hat{a}-\hat{a}^{\dag}\right), \end{equation}
where we have used the explicit form of the cavity spatial mode
function, and \begin{equation} \label{eq:U:0}
U_{0}=g_0^{2}/\Delta_{a} \end{equation} is the depth of the
single-photon dipole potential.

\subsection{Many-body dynamics} \label{Sec:Many-Body}

We now extend the previous model and derive the corresponding
effective Hamiltonian for a gas of $N$ bosons at ultralow
temperatures. The particle interactions are modeled by $s$-wave
scattering. We introduce the field operators $\Psi_j(x)$,
$\Psi_j^{\dagger}(x)$, with $j=g,e$ labeling the internal ground
state, such that \begin{eqnarray}
&&[\Psi_j(x),\Psi_i^{\dagger}(x')]=\delta_{ij}~\delta(x-x'),\\
&&[\Psi_j(x),\Psi_i(x')]=[\Psi_j^{\dagger}(x),\Psi_i^{\dagger}(x')]=0.
\end{eqnarray} In second quantization the
Hamiltonian~(\ref{eq:JC}) becomes $\mathcal{H}$ and is decomposed
according to ${\cal H}={\cal H}_0+{\cal H}_1$, where
\begin{equation}
{\cal H}_0=\sum_{j=g,e}\int\!dx\,\Psi_j^{\dag}(x)
\left(-\frac{\hbar^2\nabla^{2}}{2m}+\frac{1}{2}u_j\Psi_j^{\dag}(x)\Psi_j(x)\right)\Psi_j(x)
\end{equation} with $u_j$ the strength of the onsite interaction
depending on the atomic state, and \begin{eqnarray} {\cal H}_1
&=&-\hbar\Delta_ca^{\dagger}a-{\rm i}\hbar\eta(a-a^{\dagger})-\hbar\Delta_a\int {\rm d}x\Psi_e^{\dagger}(x)\Psi_e(x)\nonumber\\
&-&{\rm i}\hbar g_0\int {\rm
d}x\cos(kx)\left[\Psi_e^{\dagger}(x)\Psi_g(x)a-a^{\dagger}\Psi_g^{\dagger}(x)\Psi_e(x)\right].\nonumber\\
\end{eqnarray} In the above description we omit to write the
Hamiltonian term describing the collisions between atoms in
different internal states, as we will consider that the excited
state is essentially empty in the parameter regime we choose. The
quantum Heisenberg-Langevin equations for atomic and field
operators read \begin{eqnarray}
\dot{\Psi}_g(x)&=&-\frac{\rm i}{\hbar}[\Psi_g(x),{\cal H}_0]\\
& &+g_0\cos(kx)a^{\dagger}\Psi_e(x)-\sqrt{\gamma}f^{in\dagger}\Psi_e(x)\nonumber\\
\dot{\Psi}_e(x)&=&-\frac{\rm i}{\hbar}[\Psi_e(x),{\cal H}_0]+{\rm i}\Delta_a\Psi_e(x)\\
&&-g_0\cos(kx)\Psi_g(x)a-\frac{\gamma}{2}\Psi_e(x)+\sqrt{\gamma}\Psi_g(x)f^{in}\nonumber\\
\dot{a}&=&({\rm i}\Delta_c-\kappa)
a+\eta+\sqrt{2\kappa}~a^{in}\nonumber\\ &
&+g_0\int\!dx\,\cos(kx)\Psi^{\dagger}_g(x)\Psi_e(x), \label{Eq:a}
\end{eqnarray} where $f^{in}(t)$ and $a^{in}(t)$ are the noise
operators defined in the previous section. Solving the equation
for $\Psi_e(x,t)$ in the limit of large detuning,
$|\Delta_a|\gg\gamma, g_0\sqrt{\langle n\rangle}$, we find
\begin{eqnarray} \Psi_e(x)\sim -{\rm
i}\frac{g_0\cos(kx)}{\Delta_a}\Psi_g(x)a, \end{eqnarray} where the
adiabatic approximation lies on the assumption that
$\hbar|\Delta_a|\gg k_BT$, as in the single-particle case, and we
have neglected the input noise term, assuming the decay rate
$\gamma\ll|\Delta|$. Substituting this value into
Eq.~(\ref{Eq:a}), the Heisenberg-Langevin equation for the field
is given by \begin{eqnarray} \dot{a}&=&({\rm i}\Delta_c
-\kappa)a+\eta+\sqrt{2\kappa}a_{in}\nonumber-{\rm i}U_0{\mathcal
Y}a, \end{eqnarray} where \begin{equation} \label{eq:Y} {\mathcal
Y}=\int{\rm d}x\cos^2(kx)\Psi^{\dagger}_g(x)\Psi_g(x)
\end{equation} is the integral of the density of atoms in the
electronic ground state, weighted by the cavity spatial-mode function squared.
We now assume the bad-cavity limit, namely the
cavity field relaxes to the steady state on a much faster time
scale than the one in which the density of the atomic medium
varies. This limit implies $\kappa\gg k_B T/\hbar$, and
consistency with the previous assumption imposes
$|\Delta_a|\gg\kappa \gg k_B T/\hbar$. In this limit the
dependence of the field on the initial condition is negligible,
and its solution is essentially the inhomogeneous one that can be
written as \begin{eqnarray} \label{eq:A} a\simeq \eta ~F({\mathcal
Y}). \end{eqnarray} Here, we have discarded the input-noise terms,
as they are at higher order in the perturbative expansion and we
will be dealing with normally-ordered equations, so that two-time
correlations of the noise operators vanish, see
Eq.~(\ref{a:noise}). We also introduced the operator
\begin{eqnarray} F({\mathcal Y})= \frac{1}{\kappa-{\rm
i}(\Delta_c-U_0{\mathcal Y})},\nonumber\\
\label{F} \end{eqnarray} which is a function of the atom operators
in the ground state. Substituting Eq.~(\ref{eq:A}) into the equation for the
ground-state field operator we obtain \begin{eqnarray}
\label{Eq:Psi:g} \dot{\Psi}_g&=&-\frac{\rm i}{\hbar}[\Psi_g(x),{\cal
H}_0]-{\rm i}{\mathcal C}(\mathcal Y,x),
\end{eqnarray} where \begin{eqnarray} {\mathcal C}(\mathcal
Y,x)=\eta^2U_0\cos^2(kx)F^{\dagger}({\mathcal
Y})\Psi_g(x)F({\mathcal Y}).\label{C:eff}\end{eqnarray}

\subsection{Discussion} \label{Sec:Model:C}

Equation~(\ref{Eq:Psi:g}) shows explicitly the effect of the
coupling with the resonator on the atom dynamics: The coupling to
the common cavity mode induces a non-linear interaction, which
enters in the equation through operator~(\ref{F}). It is useful to
consider the average number of photons at steady state $n_{\rm
ph}=\langle a^{\dagger}a\rangle_{\rm St}$, which we obtain from
Eq.~(\ref{eq:A}) and reads \begin{equation} \label{nonlinearity}
n_{\rm ph}=\left\langle
\frac{\eta^2}{\kappa^2+(\Delta_c-U_0{\mathcal Y})^2}\right\rangle.
\end{equation} The average number of photons $n_{\rm ph}$ hence
depends on the atomic density distribution. On the other hand, it
determines the depth of the confining potential, $V\approx
\hbar|U_0|n_{\rm ph}$, and thus the atomic density distribution. In
particular, the confining potential reaches a maximum for the values
at which the denominator of Eq.~(\ref{nonlinearity}) is minimum.
From the form of operator~(\ref{eq:A}) one infers that $n_{\rm ph}$
can reach the maximum value when the parameters $\Delta_c$ and $U_0$
have the same sign (the operator $\mathcal{Y}$ is positive valued).
From Eq.~(\ref{eq:U:0}) this requires that the detunings
$\Delta_{c}$ and $\Delta_{a}$ have equal signs. This property
highlights the role of the detuning in the dynamics as control
parameters.

We now comment on the parameters required for accessing the regime
in which the effect of the non-linearity will be important, and
its consistency with the derivation we performed. We first review
the important assumptions, on which our model is based.
Spontaneous decay is neglected over the typical time scales of the
system. This imposes that the effective spontaneous scattering
rate $\gamma^{\prime}$, due to off-resonant excitation of the
dipole transition, fulfills the  inequality
$\gamma^{\prime}\ll\kappa$. Using that $\gamma^{\prime}\sim n_{\rm
ph} g_0^2\gamma/\Delta_a^2$, where $n_{\rm ph}$ is the mean value
of intracavity photons, Eq.~(\ref{nonlinearity}), then spontaneous
emission can be neglected provided that \begin{equation}
\label{decay} n_{\rm
ph}\frac{g_0^2}{\Delta_a^2}\frac{\gamma}{2}\ll \kappa.
\end{equation} As our model is based on a single-mode cavity, we
also require that the detuning between atom and cavity mode is
smaller than the free spectral range $\delta\omega$. This reduces
to the condition \begin{equation} \label{FSR} |\Delta_a|\ll
\delta\omega \end{equation} A further important assumption relies
on the relaxation time of the cavity field, which has to be much
faster than the typical time scale of atomic motion. This can be
estimated as $\kappa_BT\ll \hbar \kappa$. Finally, in order to
have that the non-linear effect on $n_{\rm ph}$ is sufficiently
large for a small number of atoms, we have required that $U_0\sim
\kappa$. This condition is however not strictly necessary: strong
nonlinear effects can be observed for smaller values of $U_0$ when the number of atoms is sufficiently large~\cite{carmichael}.

Let us now estimate the number of intracavity photons which are
usually needed, in order to find the atoms in the Mott-insulator
state. We consider specifically the case in which overlap
(bistability) regions between different Mott zone can be observed.
In Sec.~\ref{Sec:Num} we find that this occurs at values of the pump
amplitude $\eta\sim 20\kappa$. This value was evaluated for 50 to
100 atoms in the resonator. Correspondingly, the number of
intracavity photons is $n_{\rm ph}\sim 100$. From
condition~(\ref{decay}) we find that $\gamma\ll 2|\Delta|_a/n_{\rm
ph}$, where we used $U_0=\kappa$, and which is fulfilled for
$\gamma=2\pi\times 3$~MHz and $|\Delta|_a=2\pi\times 10$~GHz.
Condition~(\ref{FSR}) is then satisfied when the free spectral range
$\delta\omega$ is of the order of THz. Once $|\Delta|_a$ is fixed,
we find that $g_0\sim 2\times 0.1 \sqrt{\kappa/2\pi}$ MHz. Using the
value $\kappa=2\pi\times 53$ MHz~\cite{reichel:CQED}, this requires
$g_0\sim 2\pi \times 700$ MHz, which is presently at the border of
experimental reach.  However, for smaller values of $U_0$, say
$U_0\sim 0.1\kappa$, and for larger numbers of atoms, say $N\sim
1000$, the peculiar CQED effects on ultracold atoms we predict in
this work could be well observed for parameter regimes of present
experiments, see for instance~\cite{reichel:CQED}.

\section{The Bose-Hubbard Hamiltonian} \label{Sec:BH}

\subsection{Derivation of the Bose-Hubbard Model}
\label{Sec:BH-Derivation}

We now derive a Bose-Hubbard type of model for the dynamics of the
atoms in their self-sustained potential when the atoms are well
localized in the minima of the potential itself. Starting from the
assumption that the atoms are in a Mott-insulator state, we
decompose the atomic field operator into the operators
$b_i^{\dagger}$ and $b_i$, which create and annihilate,
respectively, atoms at the lowest band of the potential site
centered at $x=x_i$, according to \begin{equation}
\hat{\Psi}\left(x\right)=\sum_{i}\tilde{w}\left(x-x_{i}\right)\hat{b}_{i},
\end{equation} whereby $\tilde{w}(x-x_i)$ are Wannier functions,
which are to be determined by solving self-consistently the
equations of motion. The commutation rules of operators $b_i$,
$b_i^{\dagger}$ obey the bosonic commutation relations in the
regular Bose-Hubbard model, where the potential is independent of
the state of the atoms. We will show that in our case this is not
{\it apriori} warranted, due to the non-linear dependence of the
potential on the atomic density distribution, which gives rise to
a non-linear equation for the atomic wave function. However, the
bosonic commutation relations are still recovered in a properly
defined thermodynamic limit, which we will identify.

We rewrite now Eq.~(\ref{Eq:Psi:g}) within this Wannier
decomposition,  \begin{eqnarray} \label{b_ell}
\dot{b}_{\ell}=\frac{1}{{\rm i}\hbar}[b_{\ell},{\cal
H}_0^{(BH)}]-{\rm i}C, \end{eqnarray} where ${\cal H}_0^{(BH)}$
and $C$ are obtained from ${\cal H}_0$ and ${\mathcal C}$,
respectively, using the Bose-Hubbard decomposition. They read
\begin{eqnarray}\label{bh} {\cal
H}_0^{(BH)}=E_{0}\hat{N}+E_{1}\hat{B}+\frac{U}{2}\sum_{i}b_i^{\dagger}b_i^{\dagger}b_ib_i-\mu
\hat{N}, \end{eqnarray} and \begin{eqnarray}\label{C:1}
C=U_0\eta^2F^{\dagger}(\hat{Y})\left[J_0
b_{\ell}+J_1(b_{\ell+1}+b_{\ell-1})\right]F(\hat{Y}).
\end{eqnarray} The coupling matrix elements in
Eqs.~(\ref{bh})-(\ref{C:1}) read \begin{eqnarray}
E_{l}&=&\displaystyle{-\frac{\hbar^{2}}{2m}\int\!dx\,\tilde{w}\left(x-x_i\right)^{*}\nabla^{2}\tilde{w}\left(x-x_{i+l}\right)}\label{integrals1}
\\
J_{l}&=&\displaystyle{\int\!dx\,\tilde{w}\left(x-x_{i}\right)^{*}\cos^{2}\left(kx\right)\tilde{w}\left(x-x_{i+l}\right)}\label{integrals2}
\\
U&=&
\displaystyle{u_g\int\!dx\,\left|\tilde{w}\left(x\right)\right|^{4}}\label{integrals3}
\end{eqnarray} with $l=0,1$ as we keep only on-site and
nearest-neighbour couplings. In Eq.~(\ref{C:1}) we introduced the
operator \begin{eqnarray} \label{F:1}
F(\hat{Y})=\frac{1}{\kappa-{\rm i}(\Delta_c-U_0\hat{Y})},
\end{eqnarray} where operator $\hat{Y}$ is the Bose-Hubbard
decomposition of ${\mathcal Y}$, Eq.~(\ref{eq:Y}), after
neglecting couplings beyond the nearest neighbors, and takes the
form $\hat{Y}=J_0\hat{N}+J_1\hat{B}$.

In order to determine the Bose-Hubbard Hamiltonian, we now derive
an effective Hamiltonian ${\mathcal H}_{BH}$ such that
$C=[b_{\ell},{\mathcal H}_{BH}]/\hbar$. This is performed in the
limit in which we can expand operator $F$ in Eq.~(\ref{F:1}) in
the small quantity $J_1$. The details of the derivation are
reported in App.~\ref{App:A}. The Bose-Hubbard model is recovered
for a large number of atoms, according to a properly defined
thermodynamic limit. We define the thermodynamic limit by letting
$N$ and the cavity volume to infinity, keeping finite the number
of atoms per potential site. This implies the scaling $U_0\sim 1/N$.
Additionally, we impose the scaling $\eta\sim \sqrt{N}$, which corresponds to
keeping the potential depth constant as $N$ increases. This
scaling corresponds to ramping up the pump intensity with
$\sqrt{N}$. The Bose-Hubbard
type of Hamiltonian ${\mathcal H}_{\rm eff}={\mathcal
H}_0^{(BH)}+{\mathcal H}_{BH}$ is then \begin{equation}
\label{H:eff} {\mathcal H}_{\rm
eff}=E_0\hat{N}+\frac{U}{2}\sum_i\hat{n}_i(\hat{n}_i-1)-t(\hat{N})B+f(\hat{N})-\mu\hat{N},
\end{equation} where \begin{eqnarray}
&&t(\hat{N})=-E_1-\hbar\eta^2U_0J_1F^{\dagger}(J_0\hat{N})F(J_0\hat{N}) \\
&&f(\hat{N})=\frac{\hbar\eta^2}{\kappa}\arctan\left(\frac{\Delta_c-U_0J_0\hat{N}}{\kappa}\right).
\end{eqnarray} We notice that the coefficients of
Hamiltonian~(\ref{H:eff}) are operator-valued, hence imposing a
Wannier expansion such that the coefficients depend on the
operator ${\hat N}$, namely $$\tilde{w}(x-x_i)=w\left(\hat N,
x-x_{i}\right).$$ Hence, the commutation relations between the
operators $b_i$ are not the ones of bosonic operators as in the
typical Bose-Hubbard model. Nevertheless, in the thermodynamic
limit one finds \begin{equation}
[b_i,b_j^{\dagger}]=\delta_{ij}+{\mathcal O}(1/N). \end{equation}
We therefore perform the Wannier expansion in this thermodynamic
limit, consistently with the assumptions made in order to obtain
Hamiltonian~(\ref{H:eff}).

\subsection{The Bose-Hubbard Hamiltonian}

We rescale now Hamiltonian~(\ref{H:eff}) in units of the strength
of the on--site interaction $U$, which is defined in
Eq.~(\ref{integrals3}). The rescaled Hamiltonian
$\hat{\tilde{H}}=\hat{H}/U$ reads \begin{equation}\label{effham3}
\hat{\tilde{H}} =
-\tilde{t}\hat{B}+\frac{1}{2}\sum_i\hat{n}_i(\hat{n}_i-1)-\tilde{\mu}
\hat{N}, \end{equation} where \begin{equation}
\tilde{\mu}=\frac{\mu+E_0}{U}+\frac{f(\hat{N})}{\hat{N}U}
\end{equation} contains a rescaled chemical potential, while the
tunnel parameter \begin{eqnarray} \label{eq:t}
\tilde{t}&=&-\frac{E_{1}}{U}-\frac{\hbar\eta^2
U_0J_{1}}{U\left(\kappa^2+\zeta^2\right)} \end{eqnarray} is
expressed in terms of the coefficient \begin{equation}
\label{resonance} \zeta=\Delta_c-U_0J_0\hat{N}. \end{equation} The
higher order terms in $J_1\hat{B}$, describing long-range
interaction, have been neglected. Note that the number of
particles is conserved since $[\hat{N},\hat{\tilde{H}}]=0$. We
also remark that the term $f(N)/N$ tends to a constant and finite
value in the thermodynamic limit.

An important physical quantity, which will be useful for the
following study, is the depth $V$ of the cavity potential,
$V=\hbar U_0 n_{\rm ph}$ with $n_{\rm ph}$ the number of photons
in the Bose-Hubbard expansion, Eq.~(\ref{nonlinearity}). At
leading order in the expansion in $J_1$ it takes the form
\begin{equation} \label{eq:pot} V=\frac{\eta^2\hbar
U_0}{\kappa^2+\zeta^2}. \end{equation} Hamiltonian~(\ref{effham3})
and potential~(\ref{eq:pot}) are the starting points of our
analysis for the determination of the system ground state.

Let us now make some considerations about the system for a fixed
number of atoms $N$. From the form of the
potential~(\ref{eq:pot}), and in particular from the form of the
coefficient $\zeta$, Eq.~(\ref{resonance}), we observe that for
equal signs of the detunings $\Delta_a$ and $\Delta_c$ one can
have that $\zeta$ vanishes. This case corresponds to driving the
system on resonance, and gives a maximum of the cavity mode
potential. This resonance situation occurs for atom numbers $N$
that maximize the photon number, and gives rise to
bistability~\cite{carmichael}, which modifies substantially the
properties of the model with respect to the standard Bose-Hubbard one.

\section{Determination of the ground-state}
\label{Sec:GS}

In this section we determine the ground state of the system for a
fixed number of particles. Moreover, we discuss the situation when
the number of atoms is fluctuating. Our purpose is to identify the
parameter regime in which the atoms are in the Mott-insulator
state.

Starting from the assumption that the system is in the
Mott-insulator state, we use the {\it strong coupling
expansion}~\cite{monien} to verify its validity. In particular, we
apply a standard degenerate perturbative calculation in the
parameter $\tilde{t}=t/U$, and determine the ground-state energy
$E_M(n_0,\tilde{\mu},\tilde{t})$ for the Mott state with $n_0$
particles per site, and the ground-state energies
$E_{\pm}(n_0,\tilde{\mu},\tilde{t})$ when one particle is added or
subtracted to the $n_0$-th Mott state. The condition
\begin{equation} \label{Eq:Condition}
E_{M}(n_0,\tilde{\mu},\tilde{t})-E_{\pm}(n_0,\tilde{\mu},\tilde{t})=0
\end{equation} determines the boundaries $\tilde{\mu}_\pm(n_0)$ of
the $n_0$-th Mott phase as a function of the coupling parameter.
For $\tilde{\mu}_+(n_0)>\tilde{\mu}_-(n_0)$ the region between the
two chemical potentials determines the Mott zone. The Mott state
gets unstable as the parameters are varied such that
$\tilde{\mu}_+(n_0)=\tilde{\mu}_-(n_0)$ and finally
$\tilde{\mu}_+(n_0)<\tilde{\mu}_-(n_0)$. In this section we
determine the boundaries of the Mott state in a diagram, in which
we plot $\tilde{\mu}$ as a function of relevant parameters. We
remark that, in the typical Bose-Hubbard model, when the system
exits the Mott phase, then it is in a superfluid state. In our
case, this is probably verified in most cases, which we will
discuss individually.

Finally, the parameter $\tilde{t}$ can be controlled by varying
the pump amplitude $\eta$, which is straightforwardly related to
the number of of photons inside the cavity and hence to the height
of the potential. Alternatively it can be changed by varying the
atom-pump detuning $\Delta_a$ and the cavity-pump detuning
$\Delta_c$, which enter in the dynamics through the
coefficient~(\ref{resonance}) in the denominator of
Eq.~(\ref{eq:t}), and correspond to changing the refractive index
of the atomic medium.

In the following we first study the functional dependence of the
integrals on the system parameters using the Gaussian ansatz. We
then determine numerically the regions of the Mott-insulator state
in the diagram where the chemical potential $\tilde{\mu}$ is
studied as a function of the pump intensity $\eta$.

\subsection{Coefficients in the Gaussian approximation}

We determine the boundaries of the Mott-insulator regions using
the {\it Gaussian approximation}, hence replacing the Wannier
functions by Gaussian functions  in the
integrals~(\ref{integrals1})-(\ref{integrals3}). Thus, the Wannier
functions are replaced with Gaussian functions such that
\begin{equation}
\tilde{w}(x-x_i)\approx\tilde{w}_G(x-x_i)\equiv\left(\pi\sigma^2\right)^{-1/4}\mathrm{e}^{\frac{(x-x_i)^2}{2\sigma^2}},
\end{equation} where $\sigma$ is the width to be determined. This
treatment allows us to identify the dependence of the coefficients
on the physical parameters, reproducing with good approximation
the results obtained with the Wannier functions in the parameter
regimes we discuss in Sec.~\ref{Sec:Approx}. In particular, we
modify the Gaussian functions in order to fulfill the
orthogonality condition, $$\int {\rm d}x~
\tilde{w}_G'(x-x_i)\tilde{w}_G'(x-x_j)=\delta_{ij}.$$ In this way
we avoid small, but unphysical contributions.  Let $K$ be the
number of lattice sites. The width $\sigma$ of the Gaussian
functions is found from the depth $V$ of the cavity-mode
potential, Eq.~(\ref{eq:pot}). In particular,
$\sigma^2=\hbar/\sqrt{2m|V|}k$. In order to determine the
boundaries of the Mott states in the diagram of $\tilde{\mu}$ as a
function of $\eta$, we determine the coefficients for the three
cases (1) $N=Kn_0+1$, (2) $N=Kn_0$ and (3) $N=Kn_0-1$, and
introduce the subscript $(i)$ with $i=1,2,3$ for the corresponding
coefficient. We evaluate the integrals in
Eqs.~(\ref{integrals1})-(\ref{integrals3}) for these three cases
and express them as a function of the dimensionless parameter
\begin{equation}\label{eq:y:i}
y_{(i)}=k^2\sigma_{(i)}^2=\sqrt{E_R/|V_{(i)}|}, \end{equation}
where $E_R$ is the recoil energy. In term of $y_{(i)}$, they read
\begin{eqnarray} \label{coup1}
&&\displaystyle{E_{0(i)}=\frac{E_R}{2y_{(i)}}}, \\
&&\displaystyle{J_{0(i)}^\pm=\frac{1}{2}\left[1-{\rm
sign}(\Delta_a)\exp\left(-y_{(i)}\right)\right]},\label{Eq:J0} \\
&&\displaystyle{E_{1(i)}=-\frac{|V_{(i)}|}{4}\exp\left(-\frac{\pi^2}{4y_{(i)}}\right)\left(2y_{(i)}+\pi^2\right)},
\\
&&\displaystyle{J_{1(i)}^\pm=\frac{\mathrm{sign}(\Delta_a)}{2}\exp\left(-\frac{\pi^2}{4y_{(i)}}-y_{(i)}\right)},
\\
&&\displaystyle{U_{(i)}=\frac{4E_Ra_s}{\sqrt{2\pi}\Delta_{yz}}y_{(i)}},
\label{coupN} \end{eqnarray} where $a_s$ is the scattering length,
$\Delta_{yz}$ is the atomic wave packet transverse width and the
sign $\pm$ depends on the sign of $\Delta_a$. In the limit
$J_{0(i)}^\pm\gg |J_{1(i)}^\pm|$ the potential amplitude according
to (\ref{eq:pot}) is given by \begin{equation}\label{potamp}
V_{(i)}=\frac{\eta^2\hbar
U_0}{\kappa^2+\left(\Delta_c-U_0J_{0(i)}^\pm N\right)^2}.
\end{equation} As $J_{0(i)}^\pm$ depends on $V_{(i)}$ which, on
the other hand, depends itself on $J_{0(i)}^\pm$, the above
equations must be solved self-consistently. This is a consequence
of the atom-density dependence on the coupling parameters. In
particular, for $\Delta_a>0$ (atoms at the nodes), $J_{0(i)}\to 0$
in the strong pumping limit, $\eta\to\infty$, and the results
become independent of the number of atoms. On the other hand, if
$\Delta_a<0$ (atoms at the antinodes) the parameter $J_{0(i)}\to
1$ for sufficiently large pumping, and the non-linearity is
strongest.

Within this treatment we determine the nearest-neighbor coupling
parameter, which is given by \begin{equation}
\tilde{t}_{(i)}=\frac{E_R}{4\mathcal{U}}y_{(i)}^{-3/2}\mathrm{e}^{-\frac{\pi^2}{4y_{(i)}}}
\left(2y_{(i)}+\pi^2-2\mathrm{e}^{-y_{(i)}}\right), \end{equation}
where $\mathcal{U}=2\hbar^2a_sk/(\sqrt{2\pi}m\Delta_{yz})$. For
$\eta\rightarrow\infty$ the potential $|V_{(i)}|\rightarrow\infty$
and consequently $y_{(i)}\rightarrow0$, and hence
$\tilde{t}\rightarrow0$~\footnote{In the opposite limit of small
pumping we also have $\tilde{t}\rightarrow0$, which, however, is
within the regime where the Gaussian and the tight-binding
approximations are not valid}.

The perturbative calculation of the boundaries
$\tilde{\mu}_\pm(n_0)$ is sketched in the App.~\ref{App:B}. At
third order the result reads \begin{eqnarray}\label{chempot}
&&\tilde{\mu}_{+}(n_0) =
n_0+\frac{U_{(12)}}{2}Kn_0(n_0-1)-t_{(1)}2(n_0+1)\nonumber \\
&+& t_{(1)}^2n_0^2 -\left(t_{(1)}^2-t_{(2)}^2
\frac{U_{(2)}}{U_{(1)}}\right)
2Kn_0(n_0+1) \nonumber \\
&+&t_{(1)}^3n_0(n_0+1)(n_0+2), \\
&&\tilde{\mu}_{-}(n_0)  =
(n_0-1)-\frac{U_{(32)}}{2}Kn_0(n_0-1)+t_{(3)}2n_0 \nonumber \\
&-&t_{(3)}^2(n_0+1)^2+\left(t_{(3)}^2-t_{(2)}^2\frac{U_{(2)}}{U_{(3)}}\right)2Kn_0(n_0+1)
\nonumber \\
&+&t_{(3)}^3n_0(n_0+1)(n_0-1).
\end{eqnarray}
Here, $U_{(i2)}=1-U_{(2)}/U_{(i)}$,
$\tilde{\mu}_+(n_0)=\mu_+(n_0)/U_{(1)}$ and
$\tilde{\mu}_-(n_0)=\mu_-(n_0)/U_{(3)}$.

\subsection{Numerical results}
\label{Sec:Num}

In this section we study the regions of the Mott-insulator state
as a function of the chemical potential and of
the inverse pump amplitude $\eta^{-1}$. The boundaries are
determined by numerical evaluation of
Eqs.~(\ref{integrals1})-(\ref{integrals3}) using the modified
Gaussian functions. The atomic parameters we choose correspond to
$^{87}$Rb atoms with scattering length $a_s=5.77$ nm and atomic
transition wavelength $\lambda=830$ nm. The optical potential has
$K$ lattice size and the transverse width of the atomic wave
packet is $\Delta_y=\Delta_z=\sqrt{\Delta_{yz}}=30$ nm. We
evaluate the "phase diagrams" for $K=50-10000$ at fixed number of
atoms $N$, scaling $N$ so to keep the atomic density constant. The
results for the Mott zones agree over the whole range of values,
so in the figures we report the ones obtained for $K=50$ for
different values of the detunings.

\begin{figure}[ht] \begin{center}
\includegraphics[width=7cm]{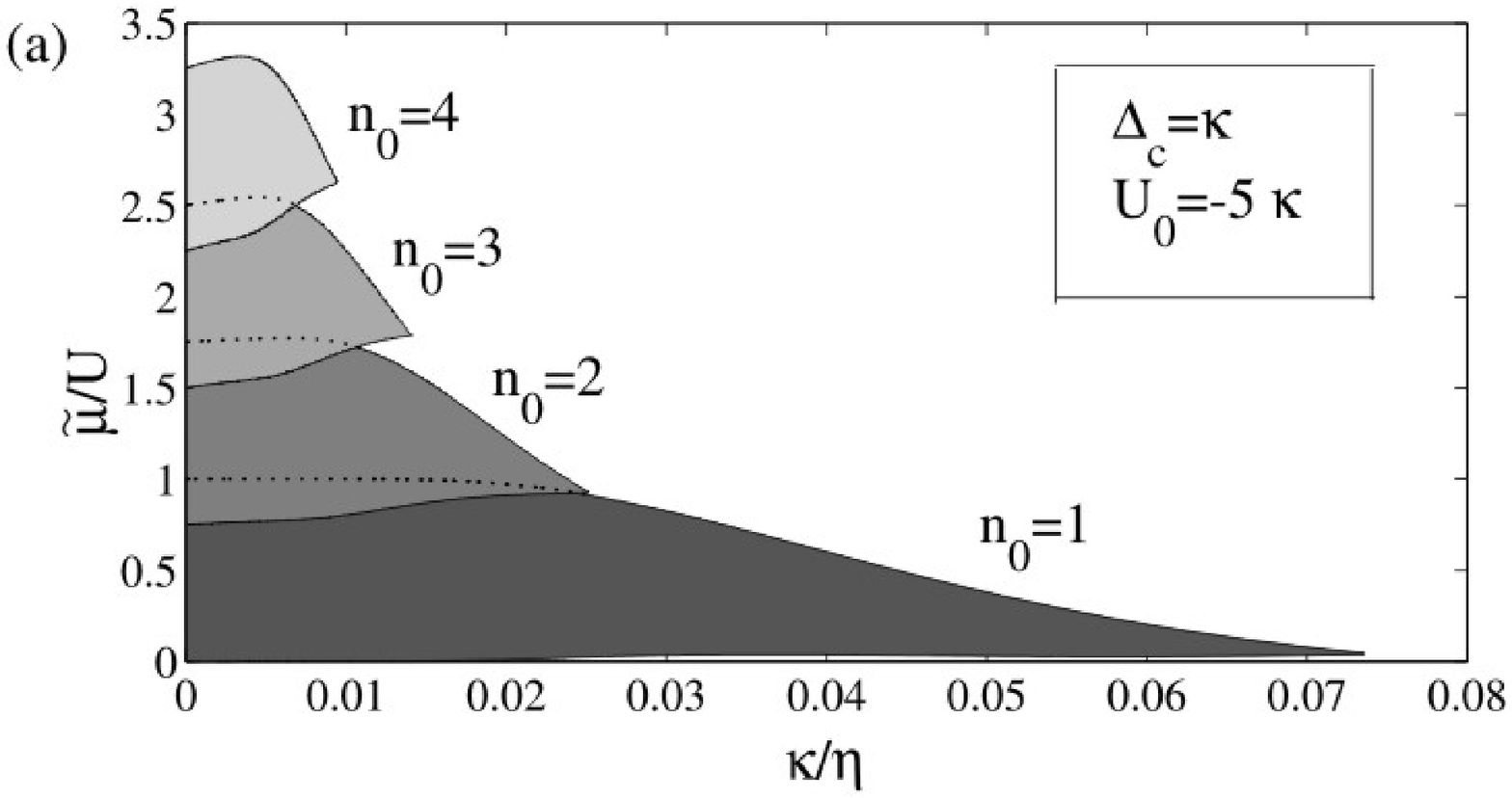}
\includegraphics[width=7cm]{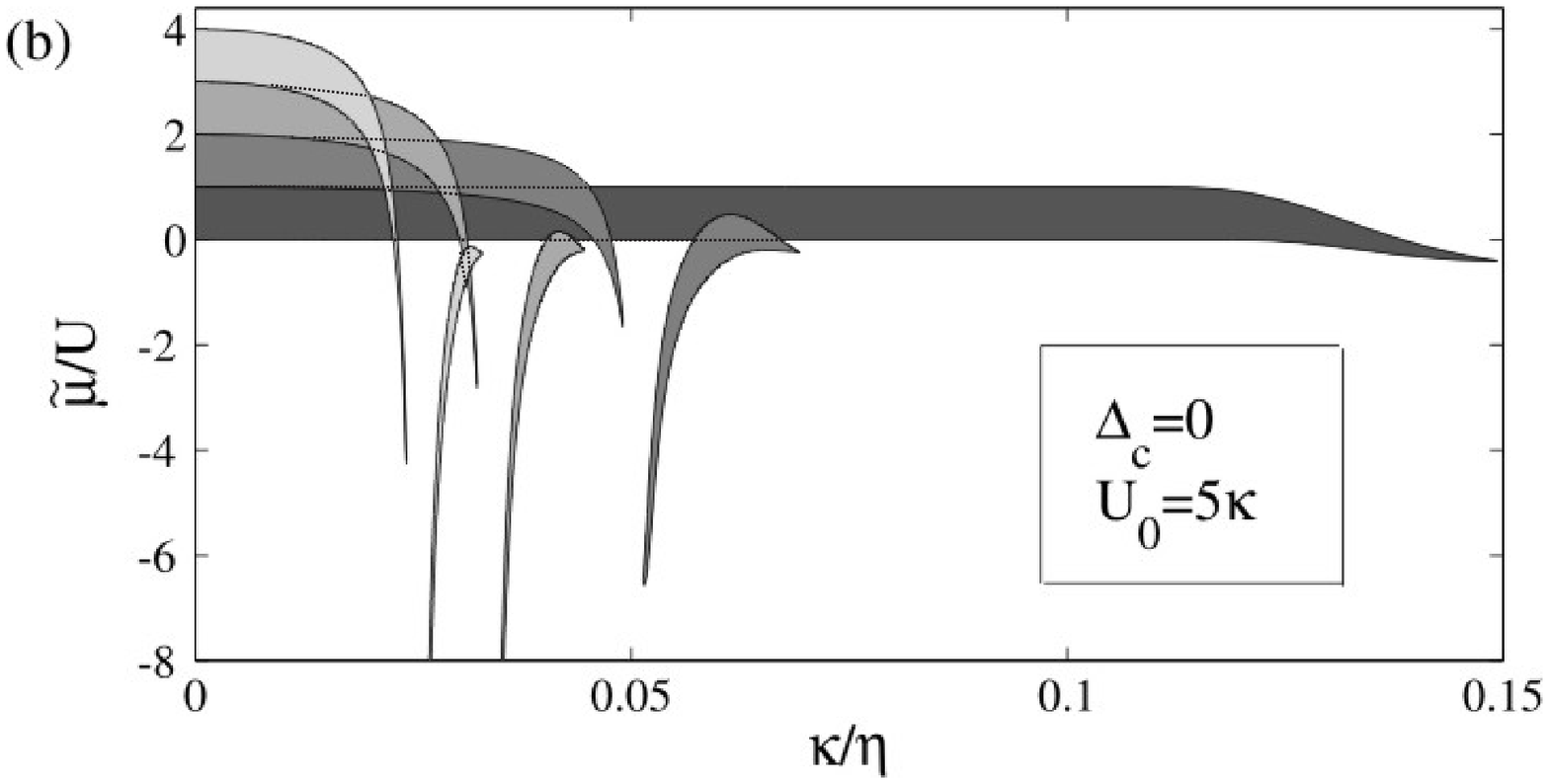}
\caption{Phase diagram, showing the Mott zones in the scaled
chemical potential-cavity pump plane $\tilde{\mu}$-$\eta^{-1}$. The
parameters are (a) $U_0=-5\kappa$ ($\Delta_a<0$, atoms at the
antinodes), $\Delta_c=2\kappa$ and (b) $U_0=5\kappa$ ($\Delta_a>0$,
atoms at the nodes), $\Delta_c=0$. The overlap and reappearance of
the Mott zones originate from the non-linearity of the system. The
dotted lines correspond to the boundaries of the covered zones. }
\label{fig1}
\end{center} \end{figure}

Figure~\ref{fig1} displays the first four Mott zones for (a)
$\Delta_a<0$ and $\Delta_c=\kappa$ and (b) $\Delta_a>0$ and
$\Delta_c=0$, as a function of the dimensionless parameter
$\kappa/\eta$. Interestingly, the extension of the Mott zones
seems to decrease roughly as $n_0^{-1}$ in both cases. We first
analyse the case displayed in Fig.~\ref{fig1}(a). For $\Delta_a<0$
the atoms are trapped at the maxima of the intracavity field.
Hence, the coupling with the cavity mode is maximum when the
confinement is very tight. Here, for large values
of the pump intensity (i.e., for small values of $\kappa/\eta$)
the Mott zones at different values of $n_0$ show some overlap.
This overlap is a cavity QED effect, in fact Mott states with
larger number of atoms per site are favored as they increase the
coupling strength to the cavity mode, and thus the depth of the
potential. The overlap is only at the border of the boundaries, as
atom-atom collisions compete with this effect. In
Fig.~\ref{fig1}(b) the detuning $\Delta_a>0$, and the atoms are
hence trapped at the nodes (the zeroes) of the intracavity field.
Hence, the coupling with the cavity mode is minimum when
$\eta\to\infty$. Indeed, here we observe that for large values of
$\eta$ (small values of $\kappa/\eta$), the Mott zones almost do
not overlap. However, for smaller values of $\eta$ they exhibit an
"exotic" behavior: overlap, disappear and reappear.

\begin{figure}[ht] \begin{center}
\includegraphics[width=6cm]{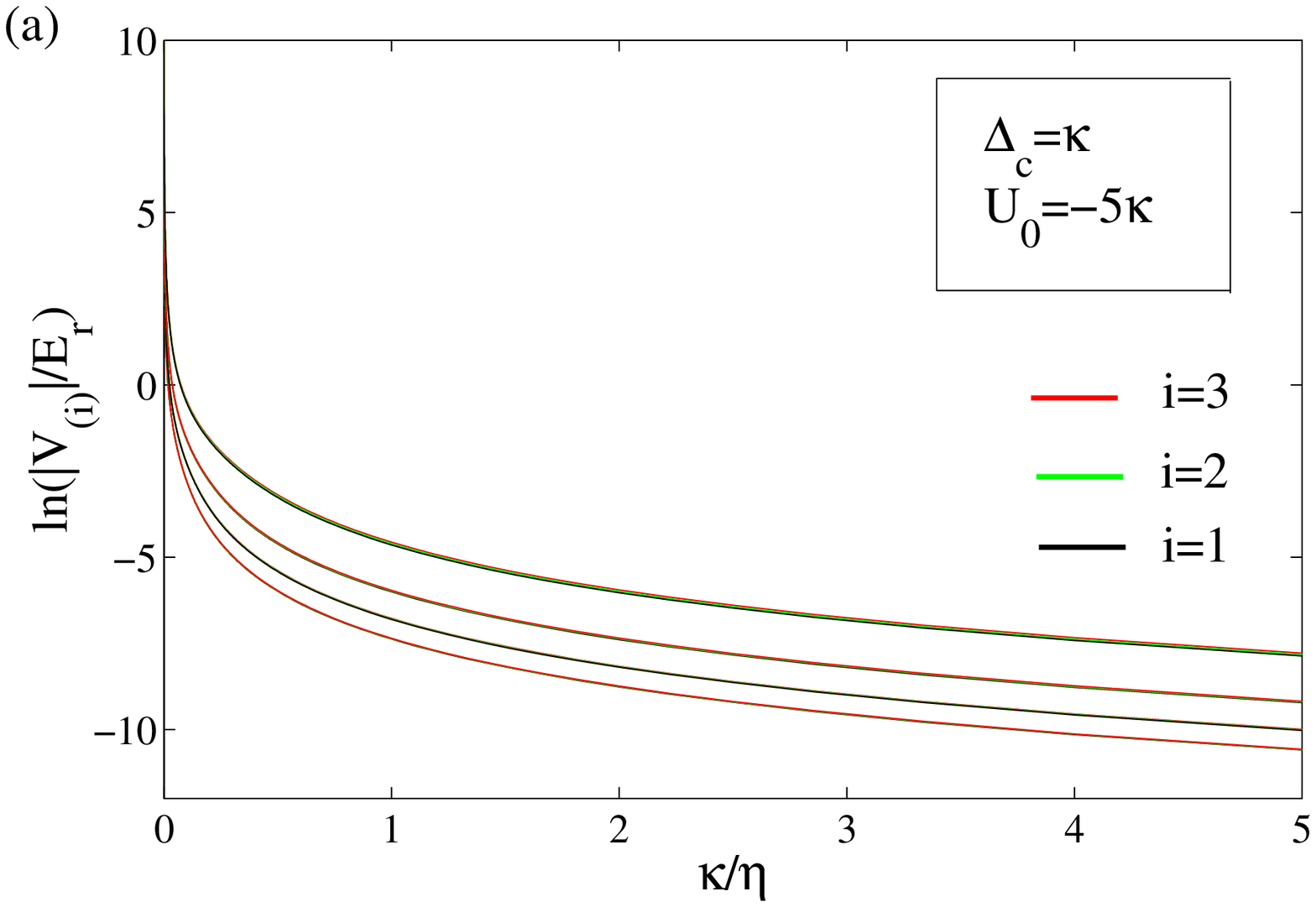}
\includegraphics[width=6cm]{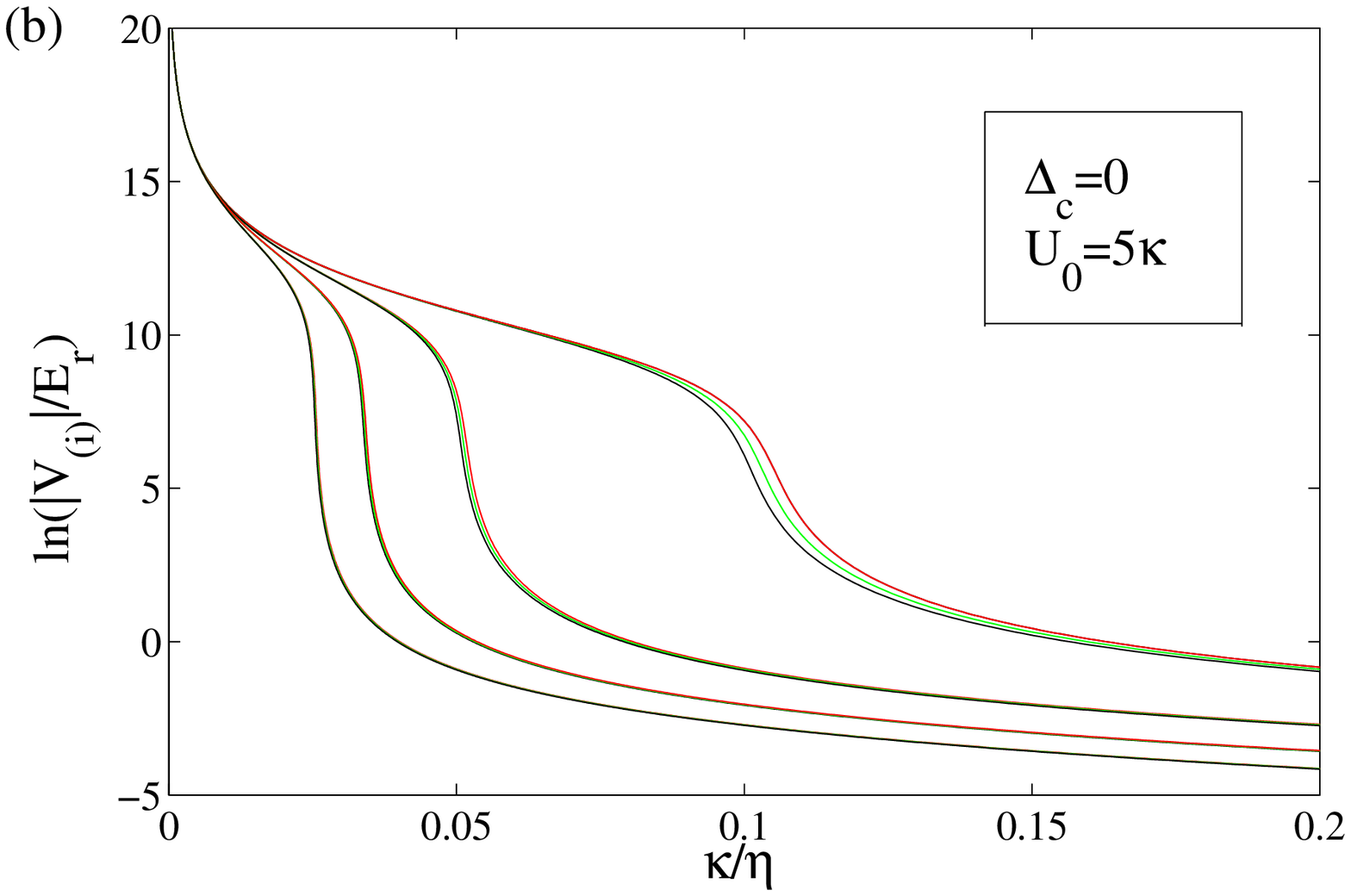}
\caption{(Colour online) The potential amplitude $|V_{(i)}|$, in
units of $E_r$ and in log-scale, as a function of the inverse
pumping $\kappa/\eta$, where the curves in~(a) and~(b) have been
evaluated in the parameter regimes of Figs~\ref{fig1}(a)
and~\ref{fig1}(b), respectively. When the non-linearity is strong
the potential amplitudes differ from the linear situation where
$|V|\sim\eta^2$. The average number of cavity photons is found by
multiplying the rescaled potential depths in the plots by the factor
$f_n=E_r/|\hbar U_0|$, which here is $f_n\approx 0.006$. }
\label{fig2} \end{center}
\end{figure}

Further insight is gained in Fig.~\ref{fig2}, where we study the
depth of the cavity potential as a function of the pump
parameters. The curves displayed in Fig.~\ref{fig2}(a) correspond
to the parameters of the phase diagram in Fig.~\ref{fig1}(a).
Here, one observes that the potential amplitude increases
monotonically as $V\sim\eta^2$ in the parameter regime where the
non-linearity is weak. Correspondingly, the width $\sigma_{(i)}$
of the Wannier functions, giving atomic localization at the minima
of the potential, decreases smoothly as
$\sigma_{(i)}\propto|V_{(i)}|^{-1/4}\sim 1/\sqrt{\eta}$, see
Eq.~(\ref{eq:y:i}). For larger pump strengths, when the
non-linearity becomes important, the behaviour is slightly
changed. The curves in Fig.~\ref{fig2}(b) correspond to the parameters
of the phase diagram in Fig.~\ref{fig1}(b). Here, one finds that
the potential depth increases rapidly where the corresponding
Mott zones exhibit a minimum in the value of $\tilde{\mu}$.
Correspondingly, the width $\sigma_{(i)}\propto|V_{(i)}|^{-1/4}$
diminishes rapidly. This behaviour changes at the value of $\eta$
where the potential gradient increases abruptly. In this regime
$\sigma_i$ varies very slowly. This can be understood as a
competition between the cavity field, which tends to localize the
atoms at the minima, and the atomic quantum fluctuations: When the
potential is sufficiently high to trap the atoms within a small
fraction of the wavelength, the cavity field is pumped more
effectively.

\begin{figure}[ht] \begin{center}
\includegraphics[width=8cm]{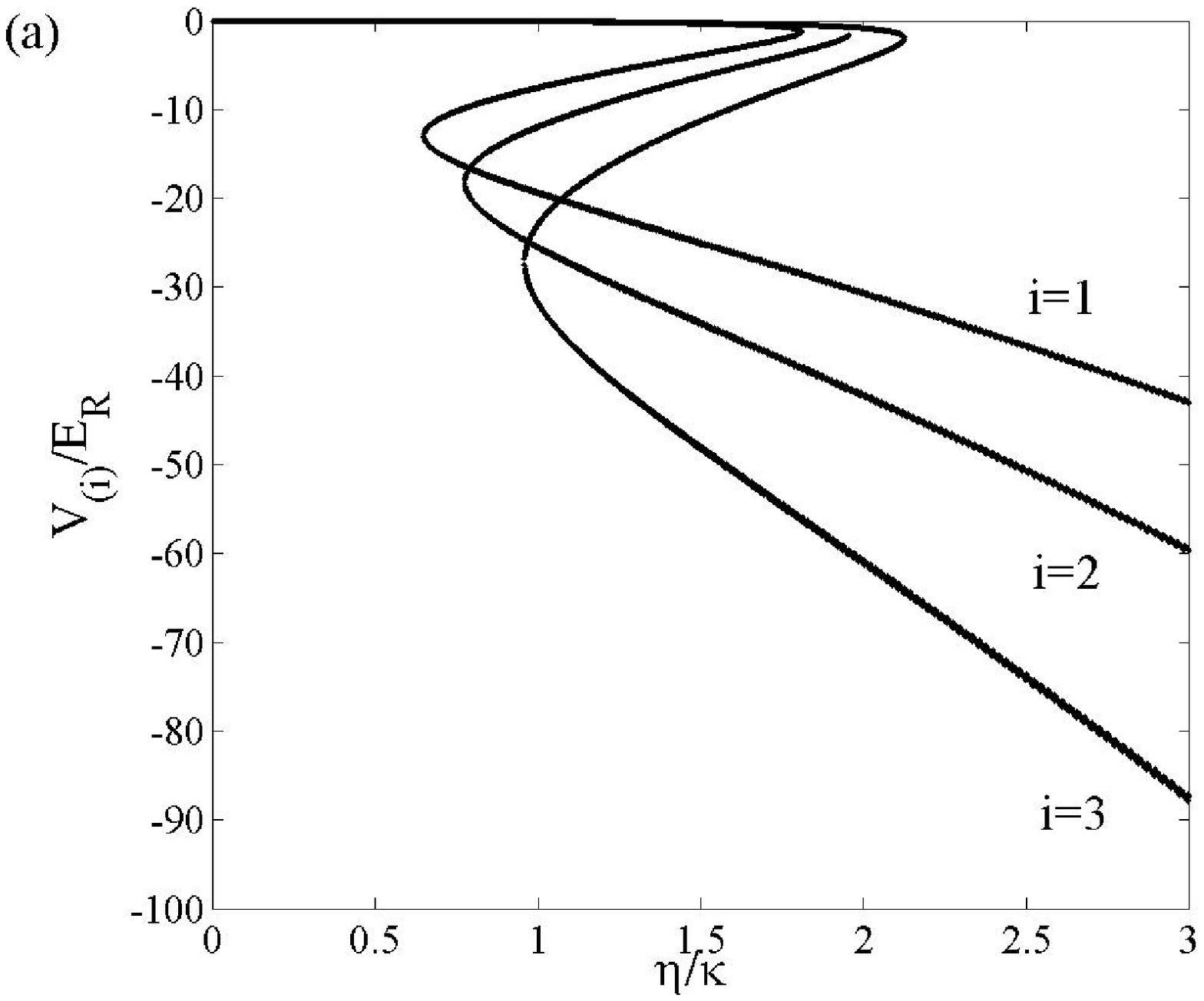}
\includegraphics[width=8cm]{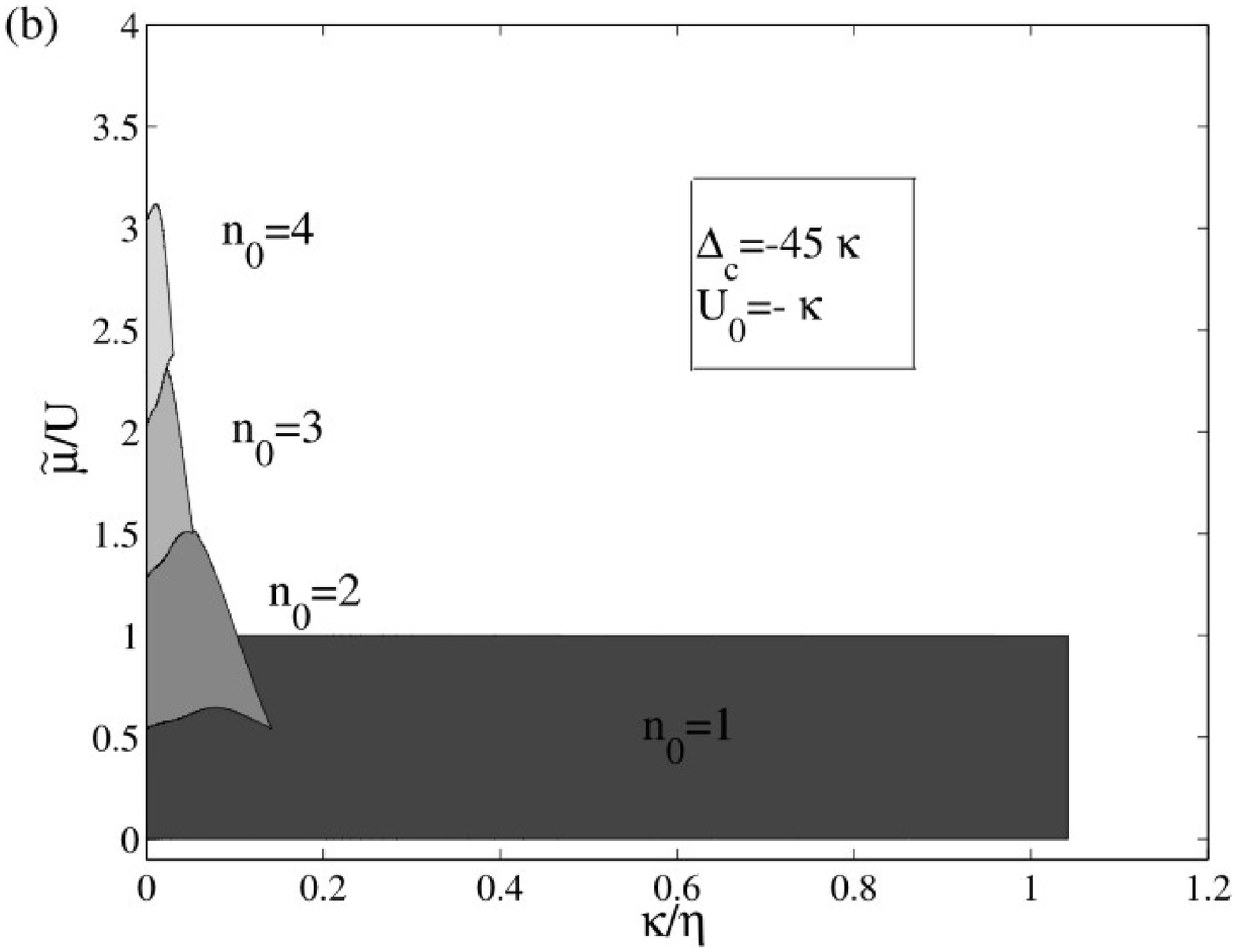} \caption{The bistability behaviour of the potential
amplitudes $V_{(i)}$ as a function of $\kappa/\eta$ (upper plot) and
corresponding phase diagram $\tilde{\mu}-\eta^{-1}$ (lower plot) for
$\Delta_c=-45\kappa$, $U_0=-\kappa$. At $n_0=1$ the Mott region
suddenly ends for $\eta\sim\kappa$, where the corresponding
potential $V_{(i)}$  jumps to a lower value. Here, the system most
likely is in a state where higher Bloch bands are populated, due to
non-adiabatic effects and the small potential depth of this
solution.} \label{fig3}
\end{center}
\end{figure}

We now consider the situation in which the detunings $\Delta_a$
and $\Delta_c$ have the same sign. In this case the parameter
$\zeta(N)$ in Eq.~(\ref{resonance}) vanishes when the condition
$J_{0(i)}^\pm=\Delta_c/U_0N$ is fulfilled, whereby
$0<J_{0(i)}^+<1/2$ and $1/2<J_{0(i)}^-<1$, see Eq.~(\ref{Eq:J0}).
This resonance condition gives rise to bistability, leading to an
abrupt change of the potential depth. As a consequence, the Mott
state may become unstable. The upper plot of Fig.~\ref{fig3}
displays the potential amplitudes $V_{(i)}/E_r$ for one atom per
site as function of $\eta/\kappa$ for $U_0=-\kappa$ and
$\Delta_c=-45\kappa$, exhibiting the typical functional behavior
of bistability. The lower plot displays the corresponding phase
diagram. For the case of one atom per site, the Mott ground-state
will suddenly disappear for $\eta\sim\kappa$ when the pumping is
adiabatically lowered. Clearly the first "jump" occurs in the
potential $V_{(i)}$ (corresponding to the lowest atom density),
and comparing the two plots one finds that this takes place
exactly when the first Mott zone suddenly ends. The system most
likely jumps into a state where higher Bloch bands are populated.
In this case the single-band and Gaussian approximations break
down.

The overlapping of the Mott zones and the bistability, which we
observe in the phase diagram, are novel features when compared
with the typical scenario of cold atomic gases trapped by an
external potential. Let us first discuss on the existence and
uniqueness of the ground-state. When the Mott-insulator state is stable,
given the number of atoms N, the ground state is fully determined
once the atomic density $\rho=N/K$ is fixed. Outside of the Mott
zones we expect superfluidity in the parameter regimes in which
there cannot be optical bistability (detunings with opposite
signs). In the situation of multiple solutions of the
Eqs.~(\ref{coup1})-(\ref{coupN}), the system will most likely be
found in the one solution which minimizes the energy.

\begin{figure}[ht] \begin{center}
\includegraphics[width=8cm]{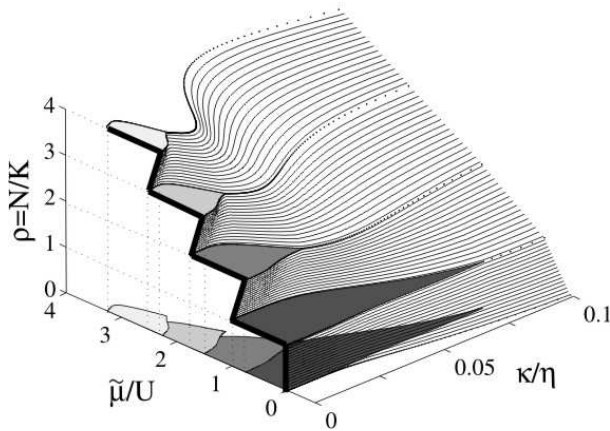}
\caption{Extended phase diagram of fig.~\ref{fig1} (a), where the
atomic density $\rho$ has been included as a third axis. Here, the
contour lines correspond to a fixed atom density $\rho$, such that
for a given $\rho$ the scaled chemical potential depends on the
pumping strength $\eta$ according to this particular contour line.
The dotted contour curves indicate the lines with exactly $n_0$
atoms per site. The projection of the Mott-zones onto the
$\tilde{\mu}-\eta^{-1}$ plane is shown.} \label{fig4} \end{center}
\end{figure}

A more complete picture of the phase diagram can be obtained by
considering the dependence on the atomic density. The strong
coupling method for higher orders is cumbersome
once the number of added/subtracted particles to the Mott states
becomes larger than one. However, the first-order corrections are
still easily obtainable for any atom number. In fig.~\ref{fig4} we
present schematically the extended phase diagram of
fig.~\ref{fig1} (a) where the atomic density has been included as
a third axis. This diagram has been obtained by fitting the
intermediate lines between the Mott zones, and by verifying that
it reproduces the first order calculations for small values of
$\kappa/\eta$. We remark that the "overhangs", corresponding to
the overlapping zones in Fig.~\ref{fig1}(a), constitute the novel
feature, which we encounter in this model as compared to the
standard Bose-Hubbard model.

\begin{figure}[ht]
\begin{center}
\includegraphics[width=8cm]{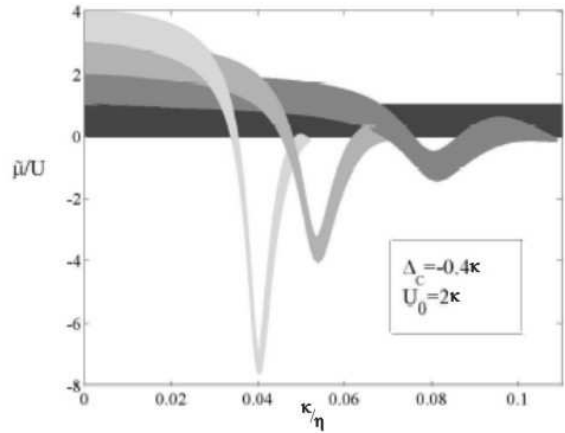}
\caption{Phase diagram on the $\tilde{\mu}-\eta^{-1}$ plane,
reporting the first four Mott zones. The Mott states with higher
energy are plotted on top of the ones with lower energy. Typically,
for large pumping the Mott zones with a large number of atoms $n_0$
per site have the highest energy. For moderate pumping this is not
necessarily true as seen by comparing for example the third and
fourth Mott zones at around $\kappa/\eta\approx0.05$. The relevant
parameters are reported in the inset.} \label{fig5}
\end{center} \end{figure}

When the number of atoms is not fixed~\cite{nonfix}, the atomic
density may take multiple values where the phase diagram exhibits
"overhangs". For sufficiently long times, we expect that the
system will be found in the number of atoms such that the energy
is minimized. This implies also that there may be a competition
between a Mott and a superfluid state at two different values of
the density, which happen to be at similar energies. Keeping this
situation in mind, we restrict our analysis to different and
overlapping Mott states, and compare their energy.
Figure~\ref{fig5} displays a phase diagram on the
$\tilde{\mu}-\eta^{-1}$ plane, whereby the Mott states with higher
energy are plotted on top of the ones with lower energy. We
observe that for large pumping strengths the Mott states with a
higher number of atoms $n_0$ have in general a greater energy,
while for lower or moderate pumping strengths this is not
necessary true. For example, the end of the Mott zone with $n_0=4$
has smaller energy than the corresponding one for the $n_0=3$. This is a pure CQED effect.

\subsection{Validity of the approximations}
\label{Sec:Approx}

We now discuss the regime of validity of the calculations, from
which we extracted the phase diagrams presented in this section.
The derivation of the system coupling parameters relies on the
assumption $J_0\gg |J_1^\pm|$. The maximum value of the ratio
$|J_1^\pm|/J_0\approx0.056$ occurs at $|V|\approx E_r/2$, hence
the nearest-neighbor coupling is at least 17 times smaller than
the on site coupling. The expansion to first order in $J_1B$ of
Eq.~(\ref{F:1}) is motivated for any number of atoms since the
perturbative parameter $\lambda\equiv J_1\hat{B}/J_0\hat{N}\sim
J_1/J_0$ is strictly smaller than unity.

The values of the chemical potential, as in Eq.~(\ref{chempot}),
are derived from a third-order perturbation expansion in the
parameter $\tilde{t}=t/U$ and it is expected to break down for
large $\tilde{t}$. We verified that in general $\tilde{t}<0.25$.
Moreover, we compared the phase diagrams with the ones obtained by
truncating at the second order in $\tilde{t}$, and could verify
that they do not differ substantially one from the others. We
remark that the perturbation calculations are carried out assuming
periodic boundary conditions, while the system here studied has a
fixed number of sites, $K=50$. We checked the validity of the
assumption by comparing the results obtained for different lattice
sites, up to 10000, keeping the density fixed.

\begin{figure}[ht]
\begin{center}
\includegraphics[width=6cm]{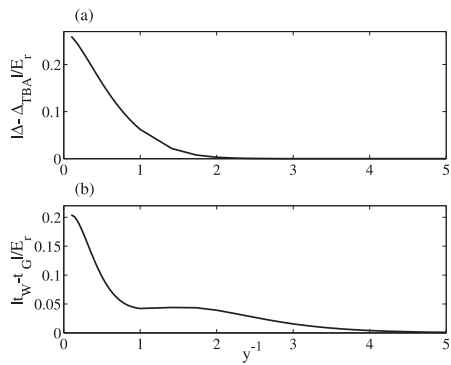}
\caption{Check of the validity of the involved
approximations. The upper figure shows the difference
$\Delta-\Delta_{TBA}$, in units of $E_R$, as a function of $y^{-1}$,
where $\Delta$ is the width of the first energy band and
$\Delta_{TBA}=2t_W$, with $t_W$ the coupling element obtained from
the corresponding Wannier functions in the tight-binding
approximation. The lower
figure displays $t_W-t_G$ as a function of $y^{-1}$, where $t_G$ is
the coupling element given by the Gaussian approximation.
The quantities are numerically derived from
Hamiltonian~(\ref{singleham}), where, in scaled
units, there is only a single parameter of interest, namely the
dimensionless potential amplitude or equivalently $y$.  }
\label{fig6}
\end{center} \end{figure}

As it concerns the tight-binding approximation (i.e., only
including nearest neighbor couplings), the single-band
approximation (i.e., expanding the field operators $\Psi(\hat{x})$
and $\Psi^\dagger(\hat{x}')$ using only the lowest band Wannier
functions), these are both related to the regime of validity of
the Gaussian approximation. Within this approximation one finds
$|J_1/J_n|=\exp\left[(n^2-1)\frac{\pi^2}{4y}\right]\gg1$, also
indicating validity of the tight-binding approximation in this
regime. Figure~\ref{fig6}(a) displays the difference
$\Delta-\Delta_{TBA}$ between the width $\Delta$ of the first
Bloch band, obtained from diagonalization of the single-particle
Hamiltonian in Eq.~(\ref{singleham}), and the width $\Delta_{TBA}$
evaluated with the tight-binding approximation, as a function of
$y^{-1}$. Figure~\ref{fig6}(b) displays the difference between the
coupling parameters obtained by using the Wannier functions and
the modified Gaussian functions as a function of $y^{-1}$. We note
that for values $y^{-1}<1$ the validity of both the tight binding
and Gaussian approximation visibly breaks down. This has also been
verified by recalculating some of the above phase-diagram using
the Wannier functions.

\section{Conclusions}
\label{Sec:Conclusions}

We have shown that ultracold bosonic atoms inside a resonator may
form stable insulator-like states, and thus enter the Mott phase,
which is sustained by and sustains the cavity potential. The low
temperature properties of the system are determined by the
competition between the quantum electrodynamic effects and the
quantum fluctuations of the atomic matter waves. This competition
gives rise to  a non-trivial dependence of the regions of
stability and of the collective atomic states on the system
parameters. Since the cavity potential depends on the state of the
atoms, the behavior of the ultracold atomic gas in the cavity
differs hence significantly from the one which is encountered in
open space. We have derived the Bose-Hubbard Hamiltonian for the
cavity confined system, and have shown that the coefficients of
this Hamiltonian depend explicitly on the number of atoms. We have
determined regions of parameters where the atomic insulator states
are stable, predicting the existence of overlapping stability
regions for competing Mott states. Bistable behavior is
encountered in the vicinity of the shifted cavity resonance,
controlled by the pump parameters.

Our theory allows us to determine the state of the atoms when their
number is fixed, while for fluctuating, non-fixed atom number, in
general, the system will choose the state of minimum energy. This
will also take place when an external inhomogeneous potential, such
as a harmonic trap potential, is additionally applied to the atoms. In
such case  we envision a possibility of hysteresis effects in the
harmonic potential, when the frequency of this potential is slowly
increased and, subsequently, slowly decreased. However, the question
how the presence of an inhomogeneous potential will modify the
insulator-like states requires further careful studies, since the
state of the system depends in a highly nontrivial way on global
parameters, which in turn determine the local density of the atoms
and the intracavity field.  The condition that atoms may affect
locally the potential, hence giving rise to phonon-like feature~\cite{Vekua},
may be reached in multi-mode resonators, allowing for localized
polaritonic excitations~\cite{Maciej,Meystre}. Further novel features are
expected when fermions are considered instead of bosons. These
questions will be tackled in future works.

\acknowledgements We thank E. Demler, T. Esslinger, Ch. Maschler, C. Menotti, H.
Monien, E. Polzik, J. Reichel, and H. Ritsch for discussions. We
acknowledge support from the Swedish
government/Vetenskapsr{\aa}det, the German DFG (SFB 407, SPP
1116), from the European Commission (EMALI, MRTN-CT-2006-035369;
SCALA, Contract No.\ 015714), from ESF PESC QUDEDIS, and the
Spanish Ministery for Education MEC (FIS 2005-04627; QLIQS,
FIS2005-08257; Ramon-y-Cajal individual fellowship; Consolider
Ingenio 2010 "QOIT").

\begin{appendix}

\section{Derivation of the effective Hamiltonian} \label{App:A}

We consider Eq.~(\ref{b_ell}), and rewrite it as \begin{equation}
\dot{b}_{\ell}=\frac{1}{{\rm i}\hbar}[b_{\ell},{\mathcal H}_0]
-{\rm i}C, \end{equation} where $C$ is defined in Eq.~(\ref{C:1})
and ${\hat Y}=J_0\hat{N}+J_1\hat{B}$. We aim at finding an
effective Hamiltonian ${\mathcal H}_{BH}$ of the Bose-Hubbard
form, such that $$C=[b_{\ell},{\mathcal H}_{BH}]/\hbar$$ in some
thermodynamic limit to be identified.

We expand now operator $C$ at first order in $J_1$, assuming
$J_1\ll J_0$, as it is verified in the Mott-insulator state, using
$[\hat{N},\hat{B}]=0$ and \begin{equation}
F(J_0\hat{N}+J_1\hat{B})\approx
F(J_0\hat{N})+J_1\hat{B}F'(J_0\hat{N}), \end{equation} where we
have introduced the notation \begin{equation}
F'(J_0\hat{N})=\frac{\partial}{\partial
y}F(y)\Bigl|_{y=J_0\hat{N}}. \end{equation} At first order in
$J_1$, we find \begin{eqnarray*} C&=&U_0\eta^2
F^{\dagger}(J_0\hat{N})[J_0
b_{\ell}+J_1(b_{\ell+1}+b_{\ell-1})]F(J_0\hat{N})\\&&+ U_0\eta^2
F^{\prime\dagger}(J_0\hat{N})J_1B J_0 b_{\ell} F(J_0\hat{N})\\&&+
U_0\eta^2F^{\dagger}(J_0\hat{N})J_0
b_{\ell}J_1BF^{\prime}(J_0\hat{N})+{\mathcal O}(J_1^2).
\end{eqnarray*} Let us now consider the commutation relations
between the various operators entering this expression. We note
that \begin{eqnarray} [b_\ell,
F(J_0\hat{N})]&=&(F(J_0(\hat{N}+1))-F(J_0\hat{N}))b_\ell\nonumber\\
&=& F'(J_0\hat{N})J_0b_\ell+{\rm O}(1/N^2) \end{eqnarray} and it
is hence of order $1/N$. Similarly, the commutator
$[b_{\ell},B]=b_{\ell+1}+b_{\ell-1}$ is at higher order in the
expansion in $1/N$. Henceforth, we can rewrite \begin{eqnarray*}
C&=&U_0\eta^2F^{\dagger}(J_0\hat{N})\left[J_0 b_{\ell}+J_1(b_{\ell+1}+b_{\ell-1})\right]F(J_0\hat{N})\\
&&+ U_0\eta^2F^{'\dagger}(J_0\hat{N}) J_0 b_{\ell} J_1
BF(J_0\hat{N})\\&&+ U_0\eta^2F^{\dagger}(J_0\hat{N})J_1B J_0
b_{\ell}F'(J_0\hat{N})\\
&\equiv&[b_{\ell},{\mathcal H}_{BH}], \end{eqnarray*} where
\begin{equation} {\mathcal H}_{BH}=\hbar
\eta^2U_0J_1F^{\dagger}(J_0\hat{N})B F(J_0\hat{N})+ G(J_0\hat
N)\end{equation} and operator $G(J_0\hat N)$ has to be determined
from the equation \begin{equation} \label{eq:G}[b_{\ell},G(J_0\hat
N)]+U_0\eta^2F^{\dagger}(J_0\hat{N})J_0 b_{\ell}F(J_0\hat{N})=0,
\end{equation} which is valid at the considered order in the
expansion in $1/N$. At leading order in $1/N$, Eq.~(\ref{eq:G}) is
a differential equation, such that $G'(J_0\hat
N)=-U_0\eta^2F^{\dagger}(J_0\hat{N})F(J_0\hat{N})$. Using the
explicit form of operator $F(x)$, Eq.~(\ref{F:1}), we find
$$G'(x)=-U_0\eta^2/(\kappa^2+(\Delta_c-U_0x)^2)$$ which gives
\begin{equation} G(x)= \frac{\eta^2}{\kappa}
\arctan\left(\frac{\Delta_c-U_0x}{\kappa}\right)\end{equation} and
finally the effective Hamiltonian in Eq.~(\ref{H:eff}).

\section{Perturbative derivation of the zone boundaries}
\label{App:B}

We consider Hamiltonian \begin{equation}\label{hamapp}
H=-\tilde{t}(\hat{N})\hat{B}+\frac{U(\hat{N})}{2}\sum_{i=1}^K\hat{n}_i(\hat{n}_i-1)-\tilde{\mu}\hat{N},
\end{equation} as given in Eq.~(\ref{effham3}). This Hamiltonian
differs from the standard Bose-Hubbard Hamiltonian, as the
coefficients depend on the operator $\hat{N}$. We apply now to
Eq.~(\ref{hamapp}) the method of Ref.~\cite{monien}, which allows
to determine the region of stability of the Mott-insulator states. The
method consists in a perturbative expansion in the parameter
$\tilde{t}$, which is assumed to be small within the parameter
regime of interest. In this limit, for large onsite interaction
strength $U$ ({\it hard core limit}), in the optical lattice the
configuration which is energetically favorable has the smallest
number of atoms per site. For a lattice of $K$ sites and
$N=Kn_0+j$ atoms, with $j<K$, there will be either $n_0$ or
$n_0+1$ atoms per site. Clearly, when $N=Kn_0$ atoms ($j=0$),
there exists only one possible ground-state, while for $N>Kn_0$
several ground state configurations exists, and one has to apply
degenerate perturbation theory.

The ground-state of Hamiltonian~(\ref{hamapp}) is found after
imposing periodic boundary conditions, and diagonalizing operator
$\hat{B}$ in the momentum representation. At $t=0$ the ground-state for $N=Kn_0$ is given by \begin{equation} \label{Psi:0}
|\Psi_0(n_0)\rangle=|n_0,n_0,...,n_0\rangle, \end{equation}
corresponding to $n_0$ atoms per site, while for $N=Kn_0+j$, with
$j>0$, they are defined by the relation
\begin{eqnarray}\label{Psi:j}
|\Psi_{j}(n_0)\rangle=\hat{A}_{k_j}^\dagger|\Psi_{j-1}(n_0)\rangle,
\end{eqnarray} where \begin{equation} \hat{A}_{k_j}^\dagger =
\frac{1}{\sqrt{K}}
\sum_{n=1}^K\mathrm{e}^{ink_ja}\frac{\hat{b}_n^\dagger}{\sqrt{\hat{n}_n+1}}
\end{equation} creates one particle in a site starting from the
lowest energy states. There is an analogous state for one hole. Here, $a=\pi/k$ is the distance between
neighboring sites, and the wave vector $k_j=2\pi j/Ka$, with
$j=-\frac{K}{2},-\frac{K}{2}+1,...,\frac{K}{2}-1$ (assuming $K$
even for simplicity). The ground state energy is calculated applying
perturbation theory in third-order in $\tilde{t}$ to this
unperturbed basis. Due to symmetry, only zeroth and second-order
in the perturbation of $t(\hat{N})\hat{B}$ contribute to the
ground-state energies of the Mott-insulator state. For $N=Kn_0$ one finds  \begin{equation}
\begin{array}{lll} E_M(n_0) & = &
\displaystyle{\frac{U_{(2)}}{2}Kn_0(n_0-1)-\mu_{(2)}Kn_0}
\\ \\
& & \displaystyle{-\frac{t_{(2)}^2}{U_{(2)}}2Kn_0(n_0+1)}
\end{array} \end{equation} while the SF energies for the added
particle/hole energies are \begin{equation} \begin{array}{lll}
E_{+}(n_0) & = &
\frac{U_{(1)}}{2}\left[Kn_0(n_0-1)+2n_0\right]-\mu_{(1)} (Kn_0+1)
\\ \\
& &
-t_{(1)}2(n_0+1)-\frac{t_{(1)}^2}{U_{(1)}}\left[2Kn_0(n_0+1)-n_0^2\right]
\\ \\
& & +\frac{t_{(1)}^3}{U_{(1)}^2}n_0(n_0+1)(n_0+2),
\\ \\
E_{-}(n_0) & = &
-\mu_{(3)}(Kn_0-1)+\frac{U_{(3)}}{2}\left[Kn_0(n_0+1)-n_0+1\right]
\\ \\
& &
-t_{(3)}2n_0-\frac{t_{(3)}^2}{U_{(3)}}\left[2Kn_0(n_0+1)-(n_0+1)^2\right]
\\ \\
& & -\frac{t_{(3)}^3}{U_{(3)}^2}n_0(n_0+1)(n_0-1). \end{array}
\end{equation} Here we have used the subscript $(i)$ corresponding
to the three different cases, $N=Kn_0$ and $N=Kn_0\pm 1$, see
Sec.~\ref{Sec:GS}. The limit of stability of the Mott-insulator
state is found when the states $|\Psi_M(n_0)\rangle$ and
$|\Psi_\pm(n_0)\rangle$ are degenerate. The conditions
$E_M(n_0)-E_+(n_0)=0$ and $E_M(n_0)-E_-(n_0)=0$ determine the
boundaries $\mu_\pm(n_0)$ of the Mott states in the phase diagram
$\tilde{\mu}-\tilde{t}$, thus obtaining the results in
Eqs.~(\ref{chempot}).

\end{appendix}

\end{document}